______________________________________________________________________________

\documentstyle[aps]{revtex}
\begin{document}

\title{Levinson's theorem for the Schr\"{o}dinger
equation in two dimensions}

\author{Shi-Hai Dong \thanks{Electronic
address:DONGSH@BEPC4.IHEP.AC.CN}}

\address{Institute of High Energy Physics, P.O.Box 918(4), Beijing
100039,
The People's Republic of China}

\author{Xi-Wen Hou}

\address{Institute of High Energy Physics,  
P.O. Box 918(4),  Beijing 100039\\
and Department of Physics,  University of Three
Gorges,  Yichang 443000,  The People's Republic of China}

\author{Zhong-Qi Ma}

\address{China Center for Advanced Science and Technology
(World Laboratory), P.O.Box 8730, Beijing 100080\\
and Institute of High Energy Physics,  P.O. Box 918(4),
Beijing 100039,  The People's Republic of China}


\maketitle

\vspace{4mm}

\begin{abstract}
Levinson's theorem for the Schr\"{o}dinger equation with a 
cylindrically symmetric potential in two dimensions is 
re-established by the Sturm-Liouville theorem. The critical 
case, where the Schr\"{o}dinger equation has a finite 
zero-energy solution, is analyzed in detail. It is shown that, 
in comparison with Levinson's theorem in non-critical case, 
the half bound state for $P$ wave, in which the wave function
for the zero-energy solution does not decay fast enough at 
infinity to be square integrable, will cause the phase shift 
of $P$ wave at zero energy to increase an additional $\pi$.

\end{abstract}

\section{INTRODUCTION}

In 1949, an important theorem in quantum mechanics was established by
Levinson [1], who set up a relation between the total number $n_{\ell}$ 
of bound states with angular momentum $\ell$ and the phase 
shift $\delta_{\ell}(0)$ of the scattering state at zero momentum
for the Schr\"{o}dinger equation with a spherically symmetric
potential $V(r)$ in three dimensions:
$$\delta_{\ell}(0)-\delta_{\ell}(\infty)=\left\{\begin{array}{ll}
\left(n_{\ell}+1/2 \right)\pi~~~~~ &{\rm when}~~\ell=0 ~~{\rm and~
a~ half~ bound~ state~ occurs} \\
n_{\ell}\pi &{\rm the~remaining~cases} , \end{array} \right.
 \eqno (1) $$

\noindent
where the potential $V(r)$ satisfies the asymptotic conditions: 
$$r^{2}|V(r)| dr \longrightarrow  0,~~~~~{\rm at}~~ r 
\longrightarrow  0, \eqno (2a) $$
$$r^{3}|V(r)| dr \longrightarrow  0,~~~~~{\rm at}~~ r
\longrightarrow \infty. \eqno(2b)$$
\noindent
The first condition is necessary for the nice behavior of 
the wave function at the origin, and the second one is 
necessary for the analytic property of the Jost function,
which was used in his proof. The first line in Eq.(1) was 
first shown by Newton [2] for the case where a half bound 
state of $S$ wave occurs. A zero-energy solution to the
Schr\"{o}dinger equation is called a half bound state if
its wave function is finite, but does not decay fast enough 
at infinity to be square integrable. As is well known, there is 
degeneracy of states for the magnetic quantum number due to 
the spherical symmetry. Usually, this degeneracy is not 
expressed explicitly in the statement of Levinson's theorem.
Due to the wide interest in lower-dimensional field theories 
recently, it may be worthwhile to study Levinson's theorem in two
dimensions. The purpose of the present paper is to re-establish
the Levinson theorem for the Schr\"{o}dinger equation in two 
dimensions in terms of the Sturm-Liouville theorem.

A lot of papers [2-10] have been devoted to the different 
proofs and generalizations of Levinson's theorem, for example,
to noncentral potentials [2], to nonlocal interactions [2,3],
to the relativistic equations [7,10], and to electron-atom
scattering [9].

Roughly speaking, there are three main methods for
the proof of Levinson's theorem. One [1] is based on 
elaborative analysis of the Jost function. This method
requires good behavior of the potential. For example,
as pointed out by Newton [11], when the asymptotic condition 
(2b) is not satisfied, Levinson's theorem is violated.
The second one is the Green function method [5], where the total
number of the physical states, which is infinite, is proved 
to be independent of the potential, and the number of the 
bound states is the difference between the infinite numbers
of the scattering states without and with the potential.
Since the number of the states in a continuous spectrum 
is uncountable, a simple model is usually used to discretize the 
continuous part of the spectrum by requiring the wave functions 
to be vanishing at a sufficiently large radius. 
We recommend the third method to prove Levinson's theorem
by the Sturm-Liouville theorem [6-8]. For the Sturm-Liouville 
problem, the fundamental trick is the definition of a phase 
angle which is monotonic with respect to the energy [12]. This 
method is very simple, intuitive and easy to generalize. In this 
proof, it is demonstrated explicitly that as the potential changes,
the phase shift at zero momentum jumps by $\pi$ while a scattering 
state becomes a bound state, or vice versa. Newton's 
counter-examples [11], where the condition (2b) is violated, can 
be proved to satisfy the modified Levinson theorem [6].

Recently, Lin [13] established a two-dimensional analog of
Levinson's theorem for the Schr\"{o}dinger equation with a 
cylindrically symmetric potential by the Green function method, 
and declared that, unlike the case in the three dimensions, 
the half bound state did not modify Levinson's theorem in two 
dimensions:
$$\eta_{m}(0)-\eta_{m}(\infty)=n_{m}\pi, ~~~~~m=0, 1, 2, \ldots,
 \eqno (3a) $$

\noindent
where $\eta_{m}(0)$ is the limit of the phase shifts at zero 
momentum for the $m$th partial wave, and $n_{m}$ is the total 
number of bound states with the angular momentum $m\hbar$. 
Both $\eta_{m}$ and $n_{m}$ are independent of the sign of
the angular momentum $\pm m\hbar$ so that only non-negative 
$m$ is needed to be discussed. The experimental study [14]
of Levinson's theorem in two dimensions has appeared in
the literatures. 

This form of Levinson's theorem for two dimensions [13] 
conflicts with an early result by Boll\'{e}, Gesztesy, 
Danneels and Wilk (BGDW) [15] in 1986, who overcame the 
difficult about the logarithmic singularity of two-dimensional 
free Green's function at zero energy, and proved with "a 
surprise" (see the title of [15]) that the half bound state 
of $P$ wave causes the phase shift $\eta_{1}(0)$ at zero momentum
to increase an additional $\pi$, exactly like the zero-energy 
bound states:
$$\eta_{m}(0)-\eta_{m}(\infty)=\left(n_{m}+1 \right)\pi, ~~~~~
{\rm when}~~m=1 ~~{\rm and~a~ half~ bound~ state~ occurs}. \eqno (3b)
$$

The critical case where the Schr\"{o}dinger equation has 
a finite zero-energy solution, is very sensitive and worthy 
of some careful analysis, especially when two conflicting 
versions of Levinson's theorem in two dimensions were presented. 
The Sturm-Liouville theorem provides a powerful tool for this 
analysis. In the present paper we re-establish the Levinson theorem
for the Schr\"{o}dinger equation in two dimensions by the
Sturm-Liouville theorem, which coincides with the version
by BGDW [15]. It seems to us that the problem in the proof 
by Lin [13] may be whether or not the set of the physical 
solutions to the Schr\"{o}dinger equation in two dimensions 
is complete when a half bound state of $P$ wave occurs 
because the corresponding wave function for the half bound
state tends to zero at infinity, although it does not decay 
fast enough at infinity to be square integrable. It
is different for $S$ wave because the wave function of 
the half bound state of $S$ wave is finite but does not 
tend to zero at infinity.

This paper is organized as follows. We firstly assume that 
the potential is vanishing beyond a sufficiently large radius 
$r_{0}$ for simplicity, and leave the discussion of the 
general potentials for the last section. In Sec.II we choose 
the logarithmic derivative of the radial wave function of the 
Schr\"{o}dinger equation as the "phase angle" [12], and prove 
by the Sturm-Liouville theorem that it is monotonic with respect 
to the energy. In terms of this monotonic property, in Sec.III 
the number of the bound states is proved to be related with the 
the logarithmic derivative of zero energy at $r_{0}$ 
as the potential changes. In Sec.IV we further prove that the 
the logarithmic derivative of zero energy at $r_{0}$ 
also determines the limit of the phase shifts at zero momentum, 
so that Levinson's theorem is proved. The critical case, where a 
zero-energy solution occurs, is analyzed carefully there. The 
problem that the potential has a tail at infinity will be 
discussed in Sec. V.

\section{NOTATIONS AND THE STURM-LIOUVILLE THEOREM}

Consider the Schr\"{o}dinger equation with a potential $V(r)$
that depends only on the distance $r$ from the origin
$$H\psi =-\displaystyle {\hbar^{2} \over 2\mu }\left(
\displaystyle {1 \over r} \displaystyle {\partial \over \partial r} 
r \displaystyle {\partial \over \partial r} + 
\displaystyle {1 \over r^{2}} \displaystyle {\partial^{2} \over 
\partial \varphi^{2} } \right)\psi +V(r) \psi =E \psi .$$ 

\noindent
where $\mu$ denotes the mass of the particle. For simplicity, we 
firstly discuss the case with a cutoff potential:
$$V(r)=0,~~~~~{\rm when}~~r\geq r_{0}, \eqno (4) $$

\noindent
where $r_{0}$ is a sufficiently large radius. The general 
case where the potential $V(r)$ has a tail at infinity will 
be discussed in Sec.V. 

Introduce a parameter $\lambda$ for the potential $V(r)$:
$$V(r,\lambda)=\lambda V(r). \eqno (5) $$

\noindent
As $\lambda$ increases from zero to one, the potential
$V(r,\lambda)$ changes from zero to the given potential
$V(r)$. 

Owing to the symmetry of the potential, we have
$$\psi(r,\varphi,\lambda)=r^{-1/2} R_{m}(r,\lambda) e^{ \pm im
\varphi},
~~~~~m=0, 1, 2, \ldots,   \eqno (6) $$  

\noindent
where the radial wave function $R_{m}(r,\lambda)$ satisfies the 
radial equation:
$$\displaystyle {\partial^{2}  R_{m}(r,\lambda) \over \partial r^{2} }
+\left\{\displaystyle {2\mu \over \hbar^{2}}
\left(E-V(r,\lambda)\right)
-\displaystyle {m^{2}-1/4 \over r^{2}} \right\} R_{m}(r,\lambda)
=0. \eqno (7) $$

Now, we are going to solve Eq.(7) in two regions and
match two solutions at $r_{0}$. Since the Schr\"{o}dinger 
equation is linear, the wave function $\psi$ can be 
multiplied by a constant factor. Removing the effect of
the factor, we only need one matching condition at $r_{0}$ for 
the logarithmic derivative of the radial function:
$$A_{m}(E,\lambda)\equiv \left\{ \displaystyle {1 \over
R_{m}(r,\lambda) }
\displaystyle {\partial R_{m}(r,\lambda) \over \partial r}
\right\}_{r=r_{0}-} 
=\left\{ \displaystyle {1 \over R_{m}(r,\lambda) }
\displaystyle {\partial R_{m}(r,\lambda) \over \partial r}
\right\}_{r=r_{0}+} . 
\eqno (8) $$

Due to the condition (2a), only one solution is convergent at the
origin. For example, for the free particle ($\lambda=0$), the solution
to Eq.(7) at the region $0\leq r \leq r_{0}$ is proportional to the 
Bessel function $J_{m}(x)$:
$$R_{m}(r,0)=\left\{\begin{array}{ll}
\sqrt{\displaystyle {\pi kr \over 2 }}J_{m}(kr),~~~~~
&{\rm when}~~E>0~~{\rm and}~~k=\left(2\mu E
\right)^{1/2}/ \hbar  \\
e^{-im\pi/2}\sqrt{\displaystyle {\pi \kappa r \over 2 }}J_{m}(i\kappa r)
,~~~~~&{\rm when}~~E\leq 0~~{\rm and}~~\kappa=\left(-2\mu E
\right)^{1/2} / \hbar , \end{array} \right. \eqno (9) $$

\noindent
The solution $R_{m}(r,0)$ given in Eq.(9) is a real function. 
A constant factor on the radial function $R_{m}(r,0)$ is not 
important.

In the region $r_{0}\leq r < \infty$, we have $V(r)=0$. For $E>0$, 
there are two oscillatory solutions to Eq.(7). Their combination 
can always satisfy the matching condition (8), so that there
is a continuous spectrum for $E>0$.
$$R_{m}(r,\lambda)=\sqrt{\displaystyle {\pi kr \over 2 }}
\left\{ \cos \eta_{m}(k,\lambda)J_{m}(kr)-\sin \eta_{m}(k,\lambda)
N_{m}(kr) \right\}~~~~~~~~~~~~~~~~~$$
$$~~~~~~~~ \sim \cos \left(kr-\displaystyle{m\pi \over 2}-
\displaystyle {\pi \over 4} +\eta_{m}(k,\lambda) \right),~~~~~~~~~~~~~
{\rm when}~~r\longrightarrow \infty . \eqno (10) $$

\noindent
where $N_{m}(kr)$ is the Neumann function.
From the matching condition (8) we have:
$$\tan \eta_{m}(k,\lambda)=\displaystyle {J_{m}(kr_{0}) \over
N_{m}(kr_{0})} ~\cdot ~\displaystyle
{A_{m}(E,\lambda)-kJ'_{m}(kr_{0})/
J_{m}(kr_{0})-1/(2r_{0})
\over A_{m}(E,\lambda)-kN'_{m}(kr_{0})/
N_{m}(kr_{0})-1/(2r_{0}) } . \eqno (11) $$
$$\eta_{m}(k)\equiv \eta_{m}(k,1). \eqno (12) $$ 

\noindent
where the prime denotes the derivative of the Bessel function, 
the Neumann function, and later the Hankel function with respect 
to their argument. 

The phase shift $\eta_{m}(k,\lambda)$ is determined from Eq.(11) 
up to a multiple of $\pi$ due to the period of the tangent 
function. Levinson determined the phase shift $\eta_{m}(k)$ 
with respect to the phase shift $\eta_{m}(\infty)$ at infinite 
momentum. For any finite potential, the phase shift 
$\eta_{m}(\infty)$ will not change and is always equal to the 
phase shift of zero potential. Therefore, Levinson's definition
for the phase shift is equivalent to the convention that the 
phase shift $\eta_{m}(k)$ is determined with respect to 
the phase shift $\eta_{m}(k,0)$ for the free particle, where 
$\eta_{m}(k,0)$ is defined to be zero:
$$\eta_{m}(k,0)=0,~~~~~{\rm where}~~V(r,0)=0. \eqno (13) $$

\noindent
We prefer to use this convention where the phase shift
$\eta_{m}(k,\lambda)$ is determined completely as $\lambda$ increases
from zero to one. It is the reason why we introduce the 
parameter $\lambda$.

Since there is only one convergent solution at infinity for $E\leq 0$ 
the matching condition (8) is not always satisfied. 
$$R_{m}(r,\lambda)=e^{i(m+1)\pi/2}\sqrt{\displaystyle 
{\pi \kappa r \over 2 }}
H^{(1)}_{m}(i\kappa r) \sim e^{-\kappa r} ,~~~~~
{\rm when}~~r\longrightarrow \infty . \eqno (14) $$

\noindent
where $H^{(1)}_{m}(x)$ is the Hankel function of the first kind.
When the condition (8) is satisfied, a bound state appears
at this energy. It means that there is a discrete spectrum 
for $E \leq 0$.

Now, we turn to the Sturm-Liouville theorem. Denote
by $\overline{R}_{m}(r,\lambda)$ the solution to Eq.(7) for the 
energy $\overline{E}$
$$\displaystyle {\partial^{2} \over \partial
r^{2}}\overline{R}_{m}(r,\lambda)
+\left\{\displaystyle {2\mu \over \hbar^{2}} \left(\overline{E}
-V(r,\lambda)\right)-\displaystyle {m^{2}-1/4 \over r^{2}} \right\} 
\overline{R}_{m}(r,\lambda)=0. \eqno (15) $$

Multiplying Eq.(7) and Eq.(15) by $\overline{R}_{m}(r,\lambda)$ and
$R_{m}(r,\lambda)$, respectively, and calculating their difference,
we have
$$\displaystyle {\partial \over \partial r} \left\{ R_{m}(r,\lambda)
\displaystyle {\partial \overline{R}_{m}(r,\lambda) \over \partial r} 
-\overline{R}_{m}(r,\lambda) \displaystyle {\partial R_{m}(r,\lambda)
\over  \partial r} \right\}
=-\displaystyle {2\mu \over \hbar^{2}}\left(\overline{E}-E\right)
\overline{R}_{m}(r,\lambda)R_{m}(r,\lambda).  \eqno (16) $$

\noindent
According to the boundary condition, both solutions $R_{m}(r,\lambda)$
and $\overline{R}_{m}(r,\lambda)$ should be vanishing at the origin. 
Integrating (16) in the region from $0$ to $r_{0}$, we have
$$\displaystyle {1 \over \overline{E}-E} \left\{ R_{m}(r,\lambda)
\displaystyle {\partial \overline{R}_{m}(r,\lambda) \over \partial r} 
-\overline{R}_{m}(r) \displaystyle {\partial R_{m}(r,\lambda) 
\over \partial r} \right\}_{r=r_{0}-}
=-\displaystyle {2\mu \over \hbar^{2}}\int_{0}^{r_{0}}
\overline{R}_{m}(r,\lambda)R_{m}(r,\lambda)dr. $$

\noindent
Taking the limit, we obtain
$$\displaystyle  {\partial A_{m}(E,\lambda) \over \partial E}
=\displaystyle  {\partial  \over \partial E}
\left( \displaystyle {1 \over R_{m}(r,\lambda) }
\displaystyle {\partial R_{m}(r,\lambda) \over \partial r}
\right)_{r=r_{0}-} 
=-\displaystyle {2\mu \over \hbar^{2}} R_{m}(r_{0},\lambda)^{-2}
\int_{0}^{r_{0}}R_{m}(r,\lambda)^{2}dr<0 . \eqno (17) $$

\noindent
Similarly, from the boundary condition that when $E\leq 0$ the radial 
function $R_{m}(r,\lambda)$ tends to zero at infinity, we have
$$\displaystyle  {\partial  \over \partial E} \left( \displaystyle 
{1 \over R_{m}(r,\lambda) }\displaystyle {\partial R_{m}(r,\lambda) 
\over \partial r} \right)_{r=r_{0}+}
=\displaystyle {2\mu \over \hbar^{2}} R_{m}(r_{0},\lambda)^{-2}
\int_{r_{0}}^{\infty}R_{m}(r,\lambda)^{2}dr>0 . \eqno (18) $$

\noindent
Therefore, when $E\leq 0$, both sides of Eq.(8) are monotonic 
with respect to the energy $E$: As energy increases, the 
logarithmic derivative of the radial function at $r_{0}-$ decreases 
monotonically, but that at $r_{0}+$ increases monotonically.
This is an essence for the Sturm-Liouville theorem. 

\section{THE NUMBER OF BOUND STATES}

In this section we will relate the number of bound states
with the logarithmic derivative $A_{m}(0,\lambda)$
of the radial function at $r_{0}-$ for zero energy when
the potential changes, in terms of the monotonic property
of the logarithmic derivative of the radial function
with respect to the energy $E$.

From Eq.(14) we have:
$$ \left( \displaystyle {1 \over R_{m}(r,\lambda) }
\displaystyle {\partial R_{m}(r,\lambda) \over \partial r}
\right)_{r=r_{0}+}
=\displaystyle {i\kappa H^{(1)}_{m}(i\kappa r_{0})' 
\over H^{(1)}_{m}(i\kappa r_{0}) }-\displaystyle {1 \over 2r_{0}}
=\left\{\begin{array}{ll} (-m+1/2)/r_{0} &{\rm when}~~E\sim 0 \\
-\kappa \sim -\infty &{\rm when}~~E\longrightarrow  -\infty.
\end{array} \right. \eqno (19) $$

\noindent
The logarithmic derivative given in Eq.(19) does not depend
on $\lambda$. On the other hand, when $\lambda=0$ we obtain 
from Eq.(10):
$$A_{m}(E,0)=\left(\displaystyle {1 \over R_{m}(r,0)}
\displaystyle {\partial R_{m}(r,0) \over \partial r }
\right)_{r=r_{0}-}
=\displaystyle {i\kappa J'_{m}(i\kappa r_{0}) 
\over J_{m}(i\kappa r_{0}) }-\displaystyle {1 \over 2r_{0}}
=\left\{\begin{array}{ll} (m+1/2)/r_{0} &{\rm when}~~E\sim 0 \\
\kappa \sim \infty &{\rm when}~~E\longrightarrow - \infty.
\end{array} \right. \eqno (20) $$

\noindent
It is evident 
from Eqs.(19) and (20) that as the energy increases from $-\infty$ 
to $0$, there is no overlap between two variant ranges of two 
logarithmic derivatives such that there is no bound state when 
$\lambda=0$ except for $S$ wave where there is a half bound 
state at $E=0$. The half bound state for $S$ wave will be 
discussed in Sec.IV.

If $A_{m}(0,\lambda)$ decreases across the value $(-m+1/2)/r_{0}$
as $\lambda$ changes, an overlap between the variant ranges 
of two logarithmic derivatives of two sides of $r_{0}$ appears. 
Since the logarithmic derivative of the radial function at $r_{0}-$ 
decreases monotonically as the energy increases, and that 
at $r_{0}+$ increases monotonically, the overlap means that
there must be one and only one energy where the matching 
condition (8) is satisfied, namely a bound state appears.
From the viewpoint of node theory, when $A_{m}(0,\lambda)$ 
decreases across the value $(-m+1/2)/r_{0}$, a node for the 
zero-energy solution to the Schr\"{o}dinger equation comes 
inwards from the infinity, namely a scattering state changes to 
a bound state.

As $\lambda$ changes, $A_{m}(0,\lambda)$ may decreases to $-\infty$, 
jumps to $\infty$, and then decreases again across the value 
$(-m+1/2)/r_{0}$, so that another overlap occurs and 
another bound state appears. Note 
that when the zero point in the zero-energy solution 
$R_{m}(r,\lambda)$ comes to $r_{0}$, $A_{m}(0,\lambda)$ goes 
to infinity. It is not a singularity. 

Each time $A_{m}(0,\lambda)$ decreases across the value
$(-m+1/2)/r_{0}$, 
a new overlap between the variant ranges of two logarithmic 
derivatives appears such that a scattering state changes to 
a bound state. In the same time, a new node comes inwards
from infinity in the zero-energy solution to the Schr\"{o}dinger 
equation. Conversely, each time $A_{m}(0,\lambda)$ increases 
across the value $(-m+1/2)/r_{0}$, an overlap between those 
two variant ranges disappears such that a bound state changes 
back to a scattering state, and simultaneously, a node goes
outwards  and disappears in the zero-energy solution. The 
number of bound states $n_{m}$ is equal to the times that 
$A_{m}(0,\lambda)$ decreases across the value $(-m+1/2)/r_{0}$ 
as $\lambda$ increases from zero to one, subtracted by the 
times that $A_{m}(0,\lambda)$ increases across the value 
$(-m+1/2)/r_{0}$. It is also equal to the number of nodes
in the zero-energy solution. 

In the next section we will show that this number is nothing but 
the phase shift $\eta_{m}(0)$ at zero momentum divided by $\pi$.

\section{LEVINSON'S THEOREM}

In order to determine the phase shift $\eta_{m}(k)$ completely,
we have introduced the convention for the phase shift 
$\eta_{m}(k,\lambda)$, where $k>0$, which is changed continuously 
as $\lambda$ increases from zero to one and $\eta_{m}(k,0)$ is 
defined to be vanishing.

The phase shift $\eta_{m}(k,\lambda)$ is calculated by Eq.(11).
It is easy to see from Eq.(11) that the phase shift 
$\eta_{m}(k,\lambda)$ increases monotonically as the 
logarithmic derivative $A_{m}(E,\lambda)$ decreases:
$$\left. \displaystyle {\partial \eta_{m}(k,\lambda) \over 
\partial A_{m}(E,\lambda)}\right|_{k}
=\displaystyle {-8r_{0}\cos^{2}\eta_{m}(k,\lambda) \over
\pi \left\{2r_{0}A_{m}(E,\lambda)N_{m}(kr_{0})-2kr_{0}N'_{m}(kr_{0})-
N_{m}(kr_{0})\right\}^{2} } \leq 0, \eqno (21) $$

\noindent
where $k=\left(2\mu E\right)^{1/2}/\hbar$.

The phase shift $\eta_{m}(0,\lambda)$ is the limit of the phase shift 
$\eta_{m}(k,\lambda)$ as $k$ tends to zero. Therefore, what we are 
interested in is the phase shift $\eta_{m}(k,\lambda)$ at a
sufficiently small momentum $k$, $k\ll 1/r_{0}$. For the small momentum
we obtain from Eq.(11)
$$\tan
\eta_{m}(k,\lambda)~~~~~~~~~~~~~~~~~~~~~~~~~~~~~~~~~~~~~~~~~~~~~~~~
~~~~~~~~~~~~~~~~~~~~~~~~~~~~~~~~~~~~~~~~~~~~~~~~~~~~~~~~~~~~~~~~~~~~~~
~~~~$$
$$=\left\{\begin{array}{ll}
\displaystyle {-\pi (kr_{0})^{2m} \over 2^{2m}m!(m-1)!}
\cdot \displaystyle {A_{m}(0,\lambda)-(m+1/2)/r_{0} \over 
A_{m}(0,\lambda)-c^{2}k^{2}-\displaystyle {-m+1/2 \over r_{0}}\left(1-
\displaystyle {(kr_{0})^{2} \over (m-1)(2m-1) }\right) } 
&{\rm when}~~m \geq 2 \\[1mm]
\displaystyle {-\pi (kr_{0})^{2} \over 4}
~\cdot~\displaystyle {A_{m}(0,\lambda)-3/(2r_{0}) \over 
A_{m}(0,\lambda)-c^{2}k^{2}+\displaystyle {1 \over 2r_{0}}
\left(1+ 2(kr_{0})^{2}\log(kr_{0})\right) } &{\rm when}~~m = 1 \\[1mm]
\displaystyle {\pi \over 2\log(kr_{0})}
~\cdot~\displaystyle {A_{m}(0,\lambda)-c^{2}k^{2}
-\displaystyle {1 \over 2r_{0}} \left(1-(kr_{0})^{2} \right) \over
A_{m}(0,\lambda)-c^{2}k^{2}-\displaystyle {1 \over 2r_{0}}\left(1+
\displaystyle {2 \over \log (kr_{0})} \right) }
 &{\rm when}~~m =0 .\end{array}\right.  \eqno (22) $$

\noindent
where the expansion for $A_{m}(E,\lambda)$, calculated from (17),
is used:
$$A_{m}(E,\lambda)=A_{m}(0,\lambda)-c^{2}k^{2}+\ldots,~~~~~c^{2}>0,~~~
~~E=\displaystyle {\hbar^{2} k^{2} \over 2\mu}. \eqno (23) $$

\noindent
In addition to the leading terms, we include in Eq.(22) some
next leading terms, which are useful only for the critical 
case where the leading terms cancel each other. 

First of all, it can be seen from Eq.(22) that 
$\tan \eta_{m}(k,\lambda)$ tends to zero as $k$ goes 
to zero, namely, $\eta_{m}(0,\lambda)$ is always equal 
to the multiple of $\pi$. In other words, if the phase 
shift $\eta_{m}(k,\lambda)$ for a sufficiently small 
$k$ is expressed as a positive or negative acute angle 
plus $n\pi$, its limit $\eta_{m}(0,\lambda)$ is equal to 
$n\pi$, where $n$ is an integer. It means that $\eta_{m}(0,\lambda)$ 
changes discontinuously. By the way, in three dimensions, the 
tangent of the phase shift may go to infinity for the critical 
case of $S$ wave.

Secondly, if $A_{m}(E,\lambda)$ decreases as $\lambda$ 
increases, $\eta_{m}(k,\lambda)$ increases monotonically. 
As $A_{m}(E,\lambda)$ decreases, each times $\tan \eta_{m}(k,\lambda)$
for a sufficiently small $k$ changes sign from positive to 
negative (through a jump from positive infinity to negative 
infinity), $\eta_{m}(0,\lambda)$ jumps by $\pi$. However, each 
times $\tan \eta_{m}(k,\lambda)$ changes sign from negative to 
positive, $\eta_{m}(0,\lambda)$ keeps invariant. Conversely, 
if $A_{m}(E,\lambda)$ increases as $\lambda$ increases, 
$\eta_{m}(k,\lambda)$ decreases monotonically. As $A_{m}(E,\lambda)$ 
increases, each time $\tan \eta_{m}(k,\lambda)$ changes sign from 
negative to positive, $\eta_{m}(0,\lambda)$ jumps by $-\pi$, and each 
time $\tan \eta_{m}(k,\lambda)$ changes sign from positive 
to negative, $\eta_{m}(0,\lambda)$ keeps invariant.  

When $V(r,\lambda)$ changes from zero to the given potential 
$V(r)$ continuously, each time the $A_{m}(0,\lambda)$ decreases
from near and larger than the value $(-m+1/2)/r_{0}$ to smaller
than that value, the denominator in Eq.(22) changes sign from
positive to negative and the remaining factor keeps positive, such that
the phase shift at zero momentum $\eta_{m}(0,\lambda)$ jumps by $\pi$.
Conversely, each time the $A_{m}(0,\lambda)$ increases across
that value, the phase shift at zero momentum $\eta_{m}(0,\lambda)$ 
jumps by $-\pi$. Note that when the $A_{m}(0,\lambda)$ decreases from 
near and larger than the value $(m+1/2)/r_{0}$ to smaller than that 
value, the numerator in Eq.(22) changes sign from positive to negative
and the remaining factor keeps negative, such that the phase shift 
at zero momentum $\eta_{m}(0,\lambda)$ does not jump. Conversely, 
when the $A_{m}(0,\lambda)$ increases across the value
$(m+1/2)/r_{0}$, 
the phase shift at zero momentum $\eta_{m}(0,\lambda)$ also keeps 
invariant. It is the reason why we did not include the
next leading terms in the numerator of Eq.(22) except for $m=0$.

Therefore, the phase shift $\eta_{m}(0)/\pi$ is just equal to the
times $A_{m}(0,\lambda)$ decreases across the value $(-m+1/2)/r_{0}$ 
as $\lambda$ increases from zero to one, subtracted by the times 
$A_{m}(0,\lambda)$ increases across that value. In the previous 
section we have proved that the difference of the two times is 
nothing but the number of bound states $n_{m}$, namely, we proved 
the Levinson theorem for the Schr\"{o}dinger equation in two 
dimensions for the non-critical cases:
$$\eta_{m}(0)=n_{m}\pi. \eqno (24a) $$

We should pay some attention to the case of $m=0$. When 
$A_{m}(0)$ decreases across the value $1/(2r_{0})$, both 
the numerator and denominator in Eq.(22) change signs, but 
not simultaneously because the next leading terms in the numerator 
and denominator of Eq.(22) are different.
It is easy to see that the numerator changes sign first, and
then the denominator changes sign, namely, $\tan \eta_{m}(k)$
at small $k$ changes firstly from negative to positive, then
to negative again so that $\eta_{m}(0)$ jumps by $\pi$.
Similarly, when $A_{m}(0)$ increases across the value
$1/(2r_{0})$, $\eta_{m}(0)$ jumps by $-\pi$.

For $\lambda=0$ ($V(r,0)=0$) and $m=0$, the numerator in Eq.(22) 
is equal to zero, the denominator is positive, and the phase shift 
$\eta_{0}(0)$ is defined to be zero. If $A_{0}(E)$ decreases 
as $\lambda$ increases from zero, the numerator becomes negative
firstly, and then the denominator changes sign from positive to negative 
such that the phase shift $\eta_{0}(0,\lambda)$ jumps by $\pi$ and 
simultaneously a bound state appears. If $A_{0}(E)$ increases 
as $\lambda$ increases from zero, the numerator becomes positive, 
and the remaining factor keeps negative such that the phase shift 
$\eta_{0}(0,\lambda)$ keeps to be zero, and no bound state 
appears. 

Now, we turn to discuss the critical case where the logarithmic 
derivative $A_{m}(0,1)$ ($\lambda=1$) is equal to the value 
$(-m+1/2)/r_{0}$. In the critical case, the following solution 
with zero energy in the region $r_{0}\leq r < \infty$ will match 
this $A_{m}(0,1)$ at $r_{0}$:
$$R_{m}(r)=r^{-m+1/2}. \eqno (25) $$

\noindent
It is a bound state when $m\geq 2$, but called a half bound state
when $m=1$ and $0$. A half bound state is not a bound state, 
because its wave function is finite but not square integrable.
We are going to discuss the critical case where $A_{m}(0,\lambda)$ 
decreases (or increases) and reaches, but not across, 
the value $(-m+1/2)/r_{0}$ as $V(r,\lambda)$ changes from zero 
to the given potential $V(r)$. For definiteness, we discuss
the case where $A_{m}(0,\lambda)$ decreases and reaches the value 
$(-m+1/2)/r_{0}$. In this case a new bound state with zero energy
appears for $m \geq 2$, but does not appear for $m=1$ and $0$. 
We should check whether or not the phase shift $\eta_{m}(0)$
increases an additional $\pi$.

It is easy to see from the next leading terms in the denominator 
of Eq.(22) that the denominator for $m \geq 2$ has changed sign 
from positive to negative as $A_{m}(0,\lambda)$ decreases and 
reaches the value $(-m+1/2)/r_{0}$, namely, the phase shift 
$\eta_{m}(0)$ jumps by $\pi$ and simultaneously a new bound 
state of zero-energy appears. 

For $m=0$ the next leading term with $\log (kr_{0})$ in the 
denominator of Eq.(22) is positive and larger than the term 
$-c^{2}k^{2}$, such that the denominator does not change sign, 
namely, the phase shift $\eta_{m}(0)$ does not jump. It meets 
the fact that no new bound state appears. 

For $m=1$ the next leading term in the denominator of Eq.(22) 
is negative such that the denominator does change sign as 
$A_{m}(0,\lambda)$ decreases and reaches the value $-1/(2r_{0})$, 
namely, the phase shift $\eta_{m}(0)$ jumps by $\pi$.
However, in this case no new bound state appears simultaneously. 

The discussion for the cases where $A_{m}(0,\lambda)$ increases and 
reaches the value $(-m+1/2)/r_{0}$ is similar. Therefore,
Levinson's theorem (24a) holds for the critical cases except 
for $m=1$. In the latter case, Levinson's theorem for the 
Schr\"{o}dinger equation in two dimensions becomes:
$$\eta_{m}(0)=\left(n_{m}+1 \right)\pi, ~~~~~
{\rm when}~~m=1 ~~{\rm and~a~ half~ bound~ state~ occurs}. \eqno (24b)
$$

\noindent
Equation (24) is the same as Eq.(3) because in our convention
$\eta_{m}(\infty)=0$.

\section{DISCUSSION}

Now, we discuss the general case where the potential $V(r)$ 
has a tail at $r\geq r_{0}$. Let $r_{0}$ be so large that only the 
leading term in $V(r)$ is concerned in the region $r\geq r_{0}$:
$$V(r) \sim \displaystyle {\hbar^{2} \over 2\mu}
br^{-n},~~~~{\rm when}~~r \longrightarrow \infty. \eqno (26) $$

\noindent
where $b$ is a nonvanishing constant and $n$ is a positive constant, 
not necessarily to be an integer. From the condition (2b), $n$ should 
be larger than 3. Substituting Eq.(26) into Eq.(7) and changing 
the variable $r$ to $\xi$
$$\xi=\left\{\begin{array}{ll}
kr=r\sqrt{2\mu E}/\hbar &{\rm when}~~E>0 \\
\kappa r=r\sqrt{-2\mu E}/\hbar &{\rm when}~~E\leq 0, \end{array}
 \right. \eqno (27) $$

\noindent
we get the radial equation at the region $r_{0} \leq r < \infty$
$$\displaystyle {d^{2} R_{m}(\xi,\lambda) \over d\xi^{2}}+
\left\{1 -\displaystyle {b \over \xi^{n}}k^{n-2} 
-\displaystyle {m^{2}-1/4 \over \xi^{2}} 
\right\} R_{m}(\xi,\lambda)=0,~~~~~{\rm when}~~E>0,$$
$$\displaystyle {d^{2} R_{m}(\xi,\lambda) \over d\xi^{2}}+
\left\{-1 -\displaystyle {b \over \xi^{n}}\kappa^{n-2} 
-\displaystyle {m^{2}-1/4 \over \xi^{2}} 
\right\} R_{m}(\xi,\lambda)=0,~~~~~{\rm when}~~E\leq 0, \eqno (28) $$

\noindent
where $R_{m}(\xi,\lambda)$ depends on $\lambda$ through the matching
condition (8).

As far as Levinson's theorem is concerned, we are only 
interested in the solutions with the sufficiently small $k$ 
and $\kappa$. If $n\geq 3$, in comparison with the term
of the centrifugal potential, the term with a factor $k^{n-2}$
(or $\kappa^{n-2}$) is too small to affect the phase 
shift at a sufficiently small $k$ and the variant range of the
logarithmic derivative $(dR_{m}(r)/dr)/R_{m}(r)$ at $r_{0}+$. 
Therefore, the proof given in the previous sections is effective
for those potential with a tail so that Levinson's theorem (24)
holds.

When $n=2$, we define
$$\nu^{2}=m^{2}+b. \eqno (29) $$ 

\noindent
The radial equation (7) becomes
$$\displaystyle {\partial^{2} R_{m}(r,\lambda) \over \partial r^{2}}+
\left\{\displaystyle {2\mu E \over \hbar^{2}}
-\displaystyle {\nu^{2}-1/4 \over r^{2}} 
\right\} R_{m}(r,\lambda)=0,~~~~~r\geq r_{0}. \eqno (30) $$

\noindent
If $\nu^{2}<0$, there are infinite number of bound states.
We will not discuss this case as well as the case with $\nu=0$ 
here. When $\nu^{2}> 0$, we take $\nu > 0$. Some formulas 
given in the previous sections will be changed by replacing 
the angular quantum number $m$ with $\nu$. Equation (19) becomes
$$ \left( \displaystyle {1 \over R_{m}(r,\lambda) }
\displaystyle {\partial R_{m}(r,\lambda) \over \partial r}
\right)_{r=r_{0}+}
=\displaystyle {i\kappa H^{(1)}_{\nu}(i\kappa r_{0})' 
\over H^{(1)}_{\nu}(i\kappa r_{0}) }-\displaystyle {1 \over 2r_{0}}
=\left\{\begin{array}{ll} (-\nu+1/2)/r_{0} &{\rm when}~~E\sim 0 \\
-\kappa \sim -\infty &{\rm when}~~E\longrightarrow  -\infty.
\end{array} \right. \eqno (31) $$

\noindent
The scattering solution (10) in the region $r_{0}\leq r < \infty$
becomes
$$R_{m}(r,\lambda)=\sqrt{\displaystyle {\pi kr \over 2 }}
\left\{ \cos \delta_{\nu}(k,\lambda)J_{\nu}(kr)
-\sin \delta_{\nu}(k,\lambda) N_{\nu}(kr)
\right\}~~~~~~~~~~~~~~~~~$$
$$~~~~~~~~ \sim \cos \left(kr-\displaystyle{\nu\pi \over 2}-
\displaystyle {\pi \over 4} +\delta_{\nu}(k,\lambda) \right)
,~~~~~~~~~~~~~
{\rm when}~~r\longrightarrow \infty . \eqno (32) $$

\noindent
Thus, the phase shift $\eta_{m}(k)$ can be calculated from
$\delta_{\nu}(k,1)$
$$\eta_{m}(k)=\delta_{\nu}(k,1)+(m-\nu)\pi/2. \eqno (33) $$

\noindent
$\delta_{\nu}(k,\lambda)$ satisfies
$$\tan \delta_{\nu}(k,\lambda)=\displaystyle {J_{\nu}(kr_{0}) \over
N_{\nu}(kr_{0})}~\cdot~ \displaystyle
{A_{m}(E,\lambda)-kJ'_{\nu}(kr_{0})/
J_{\nu}(kr_{0})-1/(2r_{0})
\over A_{m}(E,\lambda)-kN'_{\nu}(kr_{0})/
N_{\nu}(kr_{0})-1/(2r_{0}) } , \eqno (34) $$

\noindent
and it increases monotonically as the logarithmic derivative
$A_{m}(E,\lambda)$ decreases:
$$\left. \displaystyle {\partial \delta_{\nu}(k,\lambda) 
\over \partial A_{m}(E,\lambda)}\right|_{k}
=\displaystyle {-8r_{0}\cos^{2}\delta_{\nu}(k,\lambda) \over
\pi
\left\{2r_{0}A_{m}(E,\lambda)N_{\nu}(kr_{0})-2kr_{0}N'_{\nu}(kr_{0})-
N_{\nu}(kr_{0})\right\}^{2} } \leq 0. \eqno (35) $$

\noindent
For a sufficiently small $k$ we have
$$\tan
\delta_{\nu}(k,\lambda)~~~~~~~~~~~~~~~~~~~~~~~~~~~~~~~~~~~~~~~~~~~~
~~~~~~~~~~~~~~~~~~~~~~~~~~~~~~~~~~~~~~~~~~~~~~~~~~~~~~~~~~~~~~~~~~~~~~
~~~~$$
$$=\left\{\begin{array}{ll}
\displaystyle {-\pi (kr_{0})^{2\nu} \over 2^{2\nu}\nu!(\nu-1)!}
~\cdot~\displaystyle {A_{m}(0,\lambda)-(\nu+1/2)/r_{0} \over 
A_{m}(0,\lambda)-c^{2}k^{2}-\displaystyle {-\nu+1/2 \over
r_{0}}\left(1-
\displaystyle {(kr_{0})^{2} \over (\nu-1)(2\nu-1) }\right) } 
&{\rm when}~~\nu >1 \\
\displaystyle {-\pi \over \nu\Gamma(\nu)^{2}}\left(
\displaystyle  {kr_{0} \over 2}\right)^{2\nu}
\displaystyle {A_{m}(0,\lambda)-(\nu+1/2)/r_{0} \over 
A_{m}(0,\lambda)-c^{2}k^{2}-\displaystyle {-\nu+1/2 \over r_{0}}+
\displaystyle {2\pi \cot (\nu\pi) \over r_{0}\Gamma(\nu)^{2}}
\left(\displaystyle  {kr_{0} \over 2}\right)^{2\nu}  } 
&{\rm when}~~0<\nu<1  .\end{array}\right.  \eqno (36) $$

\noindent
The asymptotic forms for the case $\nu=1$ have already
been given in Eq.(22).

Now, repeating the proof for Levinson's theorem (24), we obtain
the modified Levinson's theorem for the non-critical cases:
$$\eta_{m}(0)-(m-\nu)\pi/2=\delta_{\nu}(0,1)=n_{m}\pi . \eqno (37) $$

\noindent
For the critical case where $A_{m}(0,1)=(-\nu+1/2)/r_{0}$, the
modified Levinson theorem (37) holds for $\nu>1$, where a new bound 
state appears and simultaneously $\eta_{m}(k)$ jumps by $\pi$,
but the modified Levinson theorem (37) is violated for $0<\nu \leq 1$,
where a half bound state appears and simultaneously $\eta_{m}(k)$ 
jumps by $\nu \pi$. In other words, the theorem needs to be further
modified in these cases. 

From the above discussion, we come to the conclusion that for
the potential with a tail (26) at the infinity, when $n\leq 2$
Levinson's theorem (24) is violated, and when $n>2$, even if it 
contains a logarithmic factor, Levinson's theorem (24) holds. Because
in the latter case, for any arbitrarily given small $\epsilon$,
one can always find a sufficiently large $r_{0}$ such that 
$|V(r)|<\epsilon/r^{2}$ in the region $r_{0}<r <\infty$. Since
$\nu^{2}=m^{2}+\epsilon\sim m^{2}$, Levinson's theorem (24) holds
for this case.

\vspace{6mm}
{\bf ACKNOWLEDGMENTS}. This work was supported by the National
Natural Science Foundation of China and Grant No. LWTZ-1298 of
the Chinese Academy of Sciences.

\end{document}